\documentclass[conference,a4paper]{APSIPA2021}
\usepackage{amsmath}
\usepackage{graphicx}
\usepackage{multirow}
\usepackage{threeparttable}
\usepackage[backend=biber,style=ieee,]{biblatex}
\usepackage{geometry}
\usepackage{fancyhdr}

\fancypagestyle{firststyle}{
  \fancyhf{}
  \fancyhead[C]{}
}

\usepackage[utf8]{inputenc}
\usepackage{amsfonts}
\usepackage{graphics}
\usepackage{amsmath}
\usepackage{mathrsfs}
\usepackage{xcolor}
\usepackage{subfig}
\usepackage{float}
\usepackage{acronym}
\newacro{ICE}{Independent Component Extraction}
\newacro{IVE}{Independent Vector Extraction}
\newacro{MPDR}{Minimum Power Distortionless Response}
\newacro{STFT}{Short-Time Fourier Transform}
\newacro{RMNR}{Reverberant Mixture to Noise Ratio}
\newacro{DIR}{Desired to Interference Ratio}
\newacro{DDR}{Desired to Distortion Ratio}
\newacro{DDIR}{Desired to Distortion and Interference Ratio}
\newacro{DOA}{Direction Of Arrival}
\newacro{ASR}{Automatic Speech Recognition}
\newacro{WER}{Word Error Rate}

\usepackage{booktabs}
\begin{document}

\title{Modified Parametric Multichannel Wiener Filter \\for Low-latency Enhancement of Speech Mixtures with Unknown Number of Speakers}
\author{
\authorblockN{
Ning Guo$^{1,}$\authorrefmark{1}$^,$\authorrefmark{2}, 
Tomohiro Nakatani\authorrefmark{1}, 
Shoko Araki\authorrefmark{1}, and 
Takehiro Moriya\authorrefmark{1}
}

\authorblockA{
\authorrefmark{1}
NTT Corporation, Japan \\
E-mail: tnak@ieee.org  Tel: +81-774935134}

\authorblockA{
\authorrefmark{2}
International Audio Laboratories Erlangen, Germany \\
E-mail: ning.guo@audiolabs-erlangen.de}
}

\maketitle
\thispagestyle{firststyle}
\pagestyle{empty}



\begin{abstract}
This paper introduces a novel low-latency online beamforming (BF) algorithm, named Modified Parametric Multichannel Wiener Filter (Mod-PMWF), for enhancing speech mixtures with unknown and varying number of speakers. 
%
Although conventional BFs such as linearly constrained minimum variance BF (LCMV BF) can enhance a speech mixture, they typically require such attributes of the speech mixture as the number of speakers and the acoustic transfer functions (ATFs) from the speakers to the microphones. When the mixture attributes are unavailable, estimating them by low-latency processing is challenging, hindering the application of the BFs to the problem. In this paper, we overcome this problem by modifying a conventional Parametric Multichannel Wiener Filter (PMWF). The proposed Mod-PMWF can adaptively form a directivity pattern that enhances all the speakers in the mixture without explicitly estimating these attributes. Our experiments will show the proposed BF's effectiveness in interference reduction ratios and subjective listening tests.

\end{abstract}
%
%
\section{Introduction}
\label{sec:intro}\footnotetext[1]{This work was supported by a JSPS summer program fellowship (No. SP21313)}
Microphone array processing has grown in importance due to the increasing need for remote communication. Speech signals captured by microphones often suffer from degradation caused by background noise and reverberation. Microphone array processing can improve the quality of speech signals by reducing noise and reverberation, thus enhancing speech perception and the performance of applications such as automatic speech recognition (ASR).

Researchers have successfully applied adaptive beamformers (BFs) \cite{Veen88ASSP,MPDR,ALD} to enhance a single speech signal in the captured signal by low-latency processing. Typically, they use the speech’s direction-of-arrival (DOA) to estimate the acoustic transfer function (ATF) from the speaker to the microphones (the steering vector) based on a plane-wave assumption and use the estimated ATF to optimize the BF \cite{ALD,hearingaid}. A method based on generalized eigenvalue decomposition (GEV) has been developed to estimate the ATF accurately in reverberation (i.e., multipath environments) using the spatial covariance matrices (SCMs) of the signals \cite{nonstationarity,MGolan09taslp}. Then, a Parametric Multichannel Wiener Filter (PMWF) has been proposed; It allows us to optimize a BF directly from the SCMs without estimating the ATF \cite{Souden07ASLP,ochiaie2e2017}. Furthermore, it can flexibly control the speech distortion and noise reduction tradeoff.

In contrast, enhancing a mixture of speech signals, such as a conversation, with low-latency beamforming is still challenging. This is because the mixture may contain an unknown and varying number of speakers, where their ATFs are unknown in general, and two or more speakers may talk simultaneously. In such a scenario, a BF needs to adapt its directivity pattern quickly and accurately to the unknown and varying number of speaker locations. If the number of speakers and the ATFs are given, a Linearly Constrained Minimum Variance (LCMV) BF can form a directivity pattern that enhances all the speakers in a mixture \cite{Veen88ASSP,MGolan09taslp,Er,MPLM,MCWF_Ambisonics}. Similarly, a blind source separation (BSS) technique, e.g., Independent Vector Extraction (IVE) \cite{Robin2019waspaa,ikeshita2021taslp,9616119}, can enhance a mixture by first extracting each speaker and then remixing them, when the number of speakers is known. However, the requirements by these methods on the number of speakers and their ATFs are not acceptable in the assumed scenario. Alternatively, an Unknown Reference Multichannel Wiener Filter (UR-MWF) \cite{URMWF} can enhance a mixture without knowing the number of speakers or the ATFs. However, our experiments show that it has limited denoising and dereverberation performance.


To enable low-latency speech mixture enhancement, we propose to use a modified version of PMWF, referred to as Mod-PMWF. A Mod-PMWF is defined by replacing the spatial covariance matrix (SCM) of a single target source in a conventional PMWF with the SCM of the mixture of the desired signals. Our mathematical analysis reveals that a Mod-PMWF has several desirable characteristics for low-latency speech mixture enhancement: 
\begin{enumerate}
\item A Mod-PMWF realizes a weighted sum of Minimum-Variance Distortionless Response (MVDR) BFs, each enhancing each source in the mixture. It adapts the weights of the MVDR BFs quickly to emphasize sources with larger powers at each time-frequency (TF) point. Thus, a Mod-PMWF forms a directivity pattern that enhances active sources at each TF point in the mixture.
\item We can reliably estimate a Mod-PMWF without knowing the number of sources in the mixture using a practical approximation technique. In addition, similar to a conventional PMWF, we do not need to know the individual sources' ATFs for the estimation. 
\item A Mod-PMWF has a low computational time complexity, comparable to conventional adaptive BFs such as a PMWF.
\end{enumerate}

In experiments, we pick up three conventional BFs, GEV-based MVDR (GEV-MVDR) BF \cite{MGolan09taslp}, Independent Single Component Extraction (ISCE) \cite{ikeshita2021taslp}, and UR-MWF as ones that we think may perform mixture enhancement in certain senses without knowing the number of sources or their ATFs. We also pick up Independent Multiple Component Extraction (IMCE) \cite{Robin2019waspaa} as one that can perform mixture enhancement with the prior knowledge of the number of sources. Comparison of our proposed Mod-PMWF with these conventional BFs using noisy reverberant speech mixtures confirms that the Mod-PMWF outperforms the other BFs in terms of interference reduction ratios and MUltiple Stimuli with Hidden Reference and Anchor (MUSHRA) subjective listening tests \cite{MUSHRA}.

In the following, we introduce the signal model and present the conventional PMWF in Sections~\ref{sec:sig_model} and \ref{sec:conventional_beamformer}. Section~\ref{sec:proposed_beamformers} presents the proposed Mod-PMWF, and Section~\ref{sec:conventional_methods} describes other conventional BFs that we think may be used for mixture enhancement.  Sections~\ref{sec:experiment} provide the experiment setup and results. Conclusions are drawn in Section~\ref{sec:conclusion}.

\section{Signal Model and Enhancement Goal}
\label{sec:sig_model}

\newcommand{\HT}{\mathsf{H}}

Suppose $N$ speech signals are captured by $M$ microphones with reverberation and diffuse noise. Let $x_{m,t,f}$ be the captured signal at the $m$th microphone and a TF point $(t,f)$ in the short-time Fourier transform domain for $1\le t\le T$ and $1\le f\le F$, where $T$ and $F$ are the numbers of time frames and frequency bins, and let $(\cdot)^{\top}$ denote a non-conjugate transpose. Then the captured signal at all the microphones, $\mathbf{x}_{t,f}=[x_{1,t,f},\ldots,x_{M,t,f}]^{\top}
\in\mathbb{C}^{M}$, is modeled:
\begin{align}
    \mathbf{x}_{t,f}&=\mathbf{d}_{t,f}+\mathbf{v}_{t,f},\label{eq:obs1}\\
    \mathbf{d}_{t,f}&=\sum_{n=1}^N\mathbf{d}_{t,f}^{(n)}=\sum_{n=1}^N\mathbf{h}_{f}^{(n)}s_{t,f}^{(n)},\label{eq:obs2}
\end{align}
where $\mathbf{d}_{t,f}\in\mathbb{C}^M$ is a mixture of speech signals $\mathbf{d}_{t,f}^{(n)}$ for all $n$. We assume that each speech signal, $\mathbf{d}_{t,f}^{(n)}$, comprises the direct signal plus the early reflections of the $n$th source, and can be modeled by a product of a time-invariant ATF $\mathbf{h}_{f}^{(n)}\in\mathbb{C}^{M}$ and the $n$th clean source signal $s_{t,f}^{(n)}\in\mathbb{C}$. $\mathbf{v}_{t,f}$ is the sum of all other signals, comprising the late reverberation of all sources and the additive diffuse noise. In this paper, we deal with $\mathbf{d}_{t,f}$ as the desired signal to be obtained and $\mathbf{v}_{t,f}$ as the interference signal to be reduced. In addition, we assume that $\mathbf{d}_t$ for all $n$ and $\mathbf{v}_t$ are mutually uncorrelated.


This paper assumes that both late reverberation and additive diffuse noise share the same spatial characteristics. We model their sum collectively as an interference signal $\mathbf{v}_{t,f}$ that follows stationary complex Gaussian distribution with a mean vector $\mathbf{0}$ and an SCM $\Phi_{\mathbf{v},f}\in\mathbb{C}^{M\times M}$:
\begin{align}
    p\left(\mathbf{v}_{t,f}\right)={\cal N}(\mathbf{v};\mathbf{0},\Phi_{\mathbf{v},f})\quad\mbox{where}\quad\Phi_{\mathbf{v},f}=E\{\mathbf{v}_{t,f}\mathbf{v}_{t,f}^{\HT}\},\label{eq:npdf}
\end{align}
$E\{\cdot\}$ is an expectation function, and $(\cdot)^{\HT}$ is a conjugate transpose. 

Our goal in this paper is to estimate a BF $\mathbf{W}_{t,f}\in\mathbb{C}^{M\times M}$ at each TF point that keeps the speech mixture $\mathbf{d}_{t,f}$ unchanged while minimizing $\mathbf{v}_{t,f}$. With the BF, we obtain an enhanced speech mixture, $\mathbf{y}_{t,f}\left(\approx\mathbf{d}_{t,f}\right)\in\mathbb{C}^M$:
\begin{align}
    \mathbf{y}_{t,f}=\mathbf{W}_{t,f}^{\HT}\mathbf{x}_{t,f}.
\end{align}

\section{Conventional BF for single source extraction}\label{sec:conventional_beamformer}
This section gives a brief overview of a conventional BF, PMWF, for single source extraction. It will be modified in the next section to the Mod-PMWF for mixture enhancement. 
Hereafter, we omit the index $f$ in symbols as we apply the same processing separately in each frequency.

\subsection{PMWF}
When the number of sources in the observed signal is one (i.e., $N=1$), a PMWF $\mathbf{w}_{m,t}\in\mathbb{C}^{M}$ \cite{Souden07ASLP} that enhances $\mathbf{d}_{t}^{(1)}~(=\mathbf{h}^{(1)}s_{t}^{(1)})$ in Eq.~(\ref{eq:obs2}) at the $m$th microphone is defined as one that minimizes the interference while constraining the speech distortion not exceeding an allowable level $\sigma$:
\begin{align}
    \mathbf{w}_{m,t}&=\arg\min_{\mathbf{w}} E\{|\mathbf{w}^{\HT}\mathbf{v}_t|^2\}\quad\mbox{s.t.}\quad E\{|\mathbf{w}^{\HT}\mathbf{d}_{t}^{(1)}-{d}_{m,t}^{(1)}|^2\}<\sigma.\nonumber
\end{align}
Letting $\Phi_{\mathbf{d},t}^{(1)}=E\{\mathbf{d}_{t}^{(1)}(\mathbf{d}_{t}^{(1)})^{\HT}\}$ be the SCM of $\mathbf{d}_{t}^{(1)}$ at $t$, the solution $\mathbf{W}_t=[\mathbf{w}_{1,t},\ldots,\mathbf{w}_{M,t}]$ at all microphones is obtained:
\begin{align}
\mathbf{W}_t=\left(\Phi_{\mathbf{d},t}^{(1)}+\gamma\Phi_{\mathbf{v}}\right)^{-1}\Phi_{\mathbf{d},t}^{(1)}. \label{eq:sdwmwf}
\end{align}
Here $\gamma~(\ge 0)$ is a weight controlling the tradeoff between the noise reduction and the speech distortion. This solution is also known as the Speech Distortion-Weighted MWF (SDW-MWF) \cite{SDW-MWF}. 
Then, considering that $\Phi_{\mathbf{d},t}^{(1)}$ in Eq.~(\ref{eq:sdwmwf}) is rank-1 because it can be rewritten as $\Phi_{\mathbf{d},t}^{(1)}=E\{|s_t^{(1)}|^2\}\mathbf{h}^{(1)}(\mathbf{h}^{(1)})^{\HT}$, we obtain a PMWF \cite{Souden07ASLP}:
\begin{align}
    \mathbf{W}_t^{\mbox{\scriptsize PMWF}}=\frac{\Phi_{\mathbf{v}}^{-1}\Phi_{\mathbf{d},t}^{(1)}}{\gamma+\lambda_{\mathbf{d},t}}\quad\mbox{where}\quad\lambda_{\mathbf{d},t}=\mbox{tr}\{\Phi_{\mathbf{v}}^{-1}\Phi_{\mathbf{d},t}^{(1)}\},\label{eq:orig-ss}
\end{align}
and $\mbox{tr}\{\Phi\}$ is a matrix trace. Here, we can use $\gamma$ as a parameter that controls the noise reduction level. For example, a PMWF with $\gamma=0$ becomes an MVDR BF:
\begin{align}
\mathbf{W}_t^{\mbox{\scriptsize MVDR}}=\frac{\Phi_{\mathbf{v}}^{-1}\Phi_{\mathbf{d},t}^{(1)}}{\lambda_{\mathbf{d},t}}.\label{eq:PMWF-MVDR}
\end{align}
It has also been shown that a PMWF with $\gamma=1$ becomes a MWF. 

An important advantage of a PMWF is that we can estimate it without estimating the ATF from the source to the microphones, i.e., $\mathbf{h}^{(1)}$, once we obtain $\Phi_{\mathbf{v}}$ and $\Phi_{\mathbf{d},t}^{(1)}$. Techniques to estimate $\Phi_{\mathbf{v}}$ and $\Phi_{\mathbf{d},t}^{(1)}$ have been proposed using voice activity detection \cite{Souden07ASLP} and TF masks estimated by a neural network \cite{ochiaie2e2017}.

\section{Proposed BF for mixture enhancement}\label{sec:proposed_beamformers}
This section presents the modified version of PMWF (Mod-PMWF) for mixture enhancement. We first present the definition of the Mod-PMWF, and then analyze its characteristics focusing on how it can perform mixture enhancement. Then, we describe its implementation using a practical approximation technique. 

\subsection{Definition of Mod-PMWF}
For defining the Mod-PMWF, we replace the SCM of the target signal of a PMWF, i.e., $\Phi_{\mathbf{d},t}^{(1)}$ in Eq.~(\ref{eq:orig-ss}), with the SCM of the desired mixture, i.e., $\Phi_{\mathbf{d},t}=E\{\mathbf{d}_t\mathbf{d}_t^{\HT}\}$, at each time frame. The formula of the Mod-PMWF is:
\begin{align}
    \mathbf{W}_t^{\mbox{\scriptsize Mod-PMWF}}&=\frac{\Phi_{\mathbf{v}}^{-1}\Phi_{\mathbf{d},t}}{\gamma+\lambda_{\mathbf{d},t}}\quad\mbox{where}\quad\lambda_{\mathbf{d},t}&=\mbox{tr}\{\Phi_{\mathbf{v}}^{-1}\Phi_{\mathbf{d},t}\}.\label{eq:ss-mvdr}
\end{align}
This BF is different from a PMWF in that $\Phi_{\mathbf{d},t}$ contains characteristics of not only a single target source but all the sources included in the mixture. 

In the following, we explain how a Mod-PMWF can perform mixture enhancement. Then, we present how we can reliably estimate a Mod-PMWF using an approximation in Sections~\ref{sec:approx} and \ref{sec:imple}.

\subsubsection{Analysis of Mod-PMWF}\label{sec:ana2}
Based on the assumption that $\mathbf{d}_{t}^{(n)}$ for all $n$ are mutually uncorrelated, the SCM $\Phi_{\mathbf{d},t}$ can be expanded:
\begin{align}
    \Phi_{\mathbf{d},t}=\sum_{n=1}^N\Phi_{\mathbf{d},t}^{(n)},\label{eq:scmexpand}
\end{align}
where $\Phi_{\mathbf{d},t}^{(n)}=E\{\mathbf{d}_{t}^{(n)}(\mathbf{d}_{t}^{(n)})^{\HT}\}$ is an unknown SCM of $\mathbf{d}_{t}^{(n)}$. 
Substituting Eq.~(\ref{eq:scmexpand}) into Eq.~(\ref{eq:ss-mvdr}) followed by a certain mathematical manipulation yields:
\begin{align}
    \mathbf{W}_t^{\mbox{\scriptsize Mod-PMWF}}&=\sum_{n=1}^N \mu_{n,t}\mathbf{W}_{t}^{\mbox{\scriptsize MVDR}(n)}.\label{eq:expand}
\end{align}
where $\mathbf{W}_{t}^{\mbox{\scriptsize MVDR}(n)}$ is an MVDR BF that enhances the $n$th source $\mathbf{d}_t^{(n)}$ by reducing the interference $\mathbf{v}_t$.
It is defined based on a PMWF with $\gamma=0$ in Eq.~(\ref{eq:PMWF-MVDR}) as 
\begin{align}
    \mathbf{W}_{t}^{\mbox{\scriptsize MVDR}(n)}&=\frac{\Phi_{\mathbf{v}}^{-1}\Phi_{\mathbf{d},t}^{(n)}}{\lambda_{\mathbf{d},t}^{(n)}}
    \quad\mbox{where}\quad\lambda_{\mathbf{d},t}^{(n)}=\mbox{tr}\{\Phi_{\mathbf{v}}^{-1}\Phi_{\mathbf{d},t}^{(n)}\}.
\end{align}
$\mu_{n,t}$ in Eq.~(\ref{eq:expand}) is a time-varying weight defined:
\begin{align}
    \mu_{n,t}&=\frac{\lambda_{\mathbf{d},t}^{(n)}}{\gamma+\lambda_{\mathbf{d},t}}
    \quad\mbox{where}\quad\lambda_{\mathbf{d},t}=\sum_{n=1}^N\lambda_{\mathbf{d},t}^{(n)}.\label{eq:wt1}
\end{align}

 According to Eq.~(\ref{eq:expand}), a Mod-PMWF is a weighted sum of the MVDR BFs, each of which is designed to enhance each source in the mixture. The weights are determined by $\lambda_{\mathbf{d},t}^{(n)}$, which are roughly proportional to the power of $\mathbf{d}_{t}^{(n)}$.
 
 In the following, we look into more details of the characteristics of the BF for the cases with $\gamma=0$ and $\gamma>0$.

 \subsubsection{Behavior with {$\gamma=0$}}
  When $\gamma=0$, the Mod-PMWF has the following characteristics.
 \begin{itemize}
 \item The weights sum to one according to Eqs.~(\ref{eq:wt1}), thus Eq.~(\ref{eq:expand}) becomes a weighted average of the MVDR BFs. Because all the MVDR BFs reduce $\mathbf{v}_t$, their weighted average, i.e., the Mod-PMWF, should also reduce $\mathbf{v}_t$. 
 \item MVDR BFs composing the weighted average cover all the sources in $\mathbf{d}_t$ given $\Phi_{\mathbf{d},t}$, even without explicitly specifying the number of sources.
 \item Each MVDR BF follows the unknown ATF of each source included in $\mathbf{d}_t$. Thus, it can preserve its corresponding source without distortion. 
 \item As the weights of the BFs are roughly proportional to the powers of the respective sources, the Mod-PMWF can quickly adapt the weights to focus on active speakers at each TF point, i.e., sources with larger powers.
 \end{itemize}
 In summary, each MVDR BF composing the Mod-PMWF preserves each source in the mixture while reducing the interference. The Mod-PMWF rapidly controls the weights of the MVDR BFs depending on the change in the relative powers of the sources. In this sense, the Mod-PMWF with $\gamma=0$ can achieve mixture enhancement.

\subsubsection{Behavior with {$\gamma>0$}}
With $\gamma>0$, the Mod-PMWF is equivalent to multiplying the following factor $\eta_t$ to the output of the Mod-PMWF with $\gamma=0$.
\begin{align}
\eta_t=\frac{\lambda_{\mathbf{d},t}}{\gamma+\lambda_{\mathbf{d},t}}.
\end{align}
Based on analogy from the conventional PMWF, we expect that this factor works as a single channel Wiener filter that can further reduce the interference remaining in the output of the Mod-PMWF with $\gamma=0$. Also, we confirmed such a behavior of the Mod-PMWF with $\gamma>0$ based on our preliminary experiments. More thorough analysis on this behavior should be included in future work.

\subsection{Practical approximation for Mod-PMWF}\label{sec:approx}
Although accurate estimation of $\Phi_{\mathbf{d},t}$ is crucial for Mod-PMWF, it is challenging under general recording conditions. To avoid this difficulty, this paper proposes to approximate $\Phi_{\mathbf{d},t}$ by the SCM of the captured signal $\Phi_{\mathbf{x},t}=E\{\mathbf{x}_t\mathbf{x}_t^H\}$, disregarding $\Phi_{\mathbf{v}}$ in $\Phi_{\mathbf{x},t}$, as 
\begin{align}
\Phi_{\mathbf{d},t}\simeq\Phi_{\mathbf{x},t}(=\Phi_{\mathbf{d},t}+\Phi_{\mathbf{v}}).\label{eq:approx0}
\end{align}
Unlike $\Phi_{\mathbf{d},t}$, $\Phi_{\mathbf{x},t}$ can be obtained easily and accurately from given $\mathbf{x}_t$. In addition, this approximation does not significantly affect the performance of Mod-PMWF, as discussed below. 

With this approximation, the Mod-PMWF in Eq.~(\ref{eq:ss-mvdr}) is rewritten: 
\begin{align}
    \mathbf{W}_t^{\mbox{\scriptsize Approx-Mod-PMWF}}=\frac{\Phi_{\mathbf{v}}^{-1}\Phi_{\mathbf{x},t}}{\gamma+\lambda_{\mathbf{x},t}}\quad\mbox{where}\quad\lambda_{\mathbf{x},t}=\mbox{tr}\{\Phi_{\mathbf{v}}^{-1}\Phi_{\mathbf{x},t}\}.\label{eq:approx}
\end{align}
Similar to Eq.~(\ref{eq:expand}), the approximated Mod-PMWF can be expanded using $\Phi_{\mathbf{x},t}=\Phi_{\mathbf{d},t}+\Phi_{\mathbf{v}}$: 
\begin{align}
    \mathbf{W}_t^{\mbox{\scriptsize Approx-Mod-PMWF}}=\mu_{\mathbf{d},t}'\mathbf{W}_t^{\mbox{\scriptsize Mod-PMWF}}+\mu_{\mathbf{v},t}'\frac{\mathbf{I}_M}{M},\label{eq:expand2}
\end{align}
where $\mathbf{I}_M\in\mathbb{R}^{M\times M}$ is an identity matrix. $\mu_{\mathbf{d},t}'$ and $\mu_{\mathbf{v},t}'$ are time-varying weights roughly proportional to the power of $\mathbf{d}_t$ and $\mathbf{v}_t$, and defined as
\begin{align}
\mu_{\mathbf{d},t}'=\frac{\gamma+\lambda_{\mathbf{d},t}}{\gamma+\lambda_{\mathbf{d},t}+\lambda_{\mathbf{v}}}\quad\mbox{and}\quad\mu_{\mathbf{v},t}'=\frac{\lambda_{\mathbf{v}}}{\gamma+\lambda_{\mathbf{d},t}+\lambda_{\mathbf{v}}}.
\end{align}
where $\lambda_{\mathbf{v}}=M(=\mbox{tr}\{\Phi_\mathbf{v}^{-1}\Phi_\mathbf{v}\})$. 

Equation~(\ref{eq:expand2}) is a weighted average of the Mod-PMWF (with no approximation) and a filter that reduces the gain of the captured signal by a factor of $M$. We note that $\mathbf{W}_t^{\mbox{\scriptsize Mod-PMWF}}$ in Eq.~(\ref{eq:expand2}) is accurate because it is based on accurate $\Phi_{\mathbf{d},t}$, which is included in $\Phi_{\mathbf{x},t}$.  This means that using the approximated Mod-PMWF in Eq.~(\ref{eq:approx}) is equivalent to applying an accurate Mod-PMWF and then adding the captured signal with a reduced level. Although this approximation slightly reduces the interference reduction effect of Mod-PMWF, it can precisely preserve the speech mixture. We use this approximation thoughout experiments in this paper.

\subsection{Implementation of Mod-PMWF}\label{sec:imple}
To estimate a Mod-PMWF with the approximation, we only need to estimate $\Phi_{\mathbf{x},t}$ and $\Phi_{\mathbf{v},t}$ and calculate Eq.~(\ref{eq:approx}) at each TF point. 
First, according to the convention of low-latency online adaptive BFs, we can estimate $\Phi_{\mathbf{x},t}$ by online processing as the time average of $\mathbf{x}_t\mathbf{x}_t^{\HT}$ using a forgetting factor $\beta$ ($0<\beta\le 1)$. It can be recursively updated:
\begin{align}
    \Phi_{\mathbf{x},t}=\beta\Phi_{\mathbf{x},t-1}+(1-\beta)\mathbf{x}_t\mathbf{x}_t^{\HT}.\label{eq:scmupdate}
\end{align}
Note that active sources may differ at different time frames, and sources that are not included in the current frame $t$ may be included in the past captured signals in Eq.~(\ref{eq:scmupdate}). Then, their influence remains in $\Phi_{\mathbf{x},t-1}$. To minimize such influence, it is desirable to use a small $\beta$.

As for $\Phi_{\mathbf{v},t}$, $\Phi_{\mathbf{v},t}$ can be estimated, e.g., from noise signals recorded separately in advance, or from the captured signal during speech absent periods. Also, we can use the same techniques proposed for conventional PMWF \cite{Souden07ASLP,ochiaie2e2017}. 

\section{Conventional BFs that may be applicable to mixture enhancement}\label{sec:conventional_methods}
This section describes several conventional BFs that we think may be used for mixture enhancement in certain senses. We also give a comparison of the computational time complexity of the BFs with our proposed Mod-PMWF. In the next section, we will compare the BFs with our proposed Mod-PMWF by experiments.

\subsection{GEV-MVDR BF}
The first conventional BF is a GEV-MVDR BF. We first present it as a technique for single source enhancement, and then explain how we can use it for mixture enhancement. 

When the number of sources in the observed signal is one (i.e., $N=1$), an ATF at each time frame $\mathbf{h}^{(1)}$ for the source can be estimated based on GEV \cite{MGolan09taslp}:
\begin{align}
    \mathbf{h}_t^{(1)}=\Phi_{\mathbf{v}}\mathbf{u}_t\quad\mbox{where}\quad\mathbf{u}_t=\mbox{MaxEig}\{\Phi_{\mathbf{v}}^{-1}\Phi_{\mathbf{x},t}\},\label{eq:maxeig}
\end{align}
where $\mbox{MaxEig}\{\Phi\}$ extracts an eigenvector of $\Phi$ corresponding to the maximum eigenvalue. 
Considering $\Phi_{\mathbf{d},t}^{(1)}=E\{|s_t^{(1)}|^2\}\mathbf{h}_t^{(1)}(\mathbf{h}_t^{(1)})^{\HT}$ in Eq.~(\ref{eq:PMWF-MVDR}) and substituting Eq.~(\ref{eq:maxeig}) into Eq.~(\ref{eq:PMWF-MVDR}), we obtain a GEV-MVDR BF:
\begin{align}
    \mathbf{W}_t^{\mbox{\scriptsize GEV-MVDR}}&=\frac{\mathbf{u}_{t}\mathbf{u}_{t}^{\HT}\Phi_{\mathbf{v}}}{\mbox{tr}\{\mathbf{u}_{t}\mathbf{u}_{t}^{\HT}\Phi_{\mathbf{v}}\}}.\label{eq:gev-mvdr}
\end{align}

For low-latency processing, we can calculate $\Phi_{\mathbf{x},t}$ in Eq.~(\ref{eq:maxeig}) using Eq.~(\ref{eq:scmupdate}). Then, we can obtain $\mathbf{u}_t$ in Eq.~(\ref{eq:gev-mvdr}) in a computationally efficient way using a power method for calculating $\mbox{MaxEig}\{\cdot\}$ in Eq.~(\ref{eq:maxeig}) \cite{ATFtracking,nakatani19_interspeech}. This iterates the following update after first setting $\mathbf{u}_t=\mathbf{u}_{t-1}$ at each time frame:
\begin{align}
    \mathbf{u}_t\leftarrow\Phi_{\mathbf{v}}^{-1}\Phi_{\mathbf{x},t}\mathbf{u}_{t}/|\mathbf{u}_{t}|,
\end{align}
In general, the power method converges very quickly, and only one iteration was sufficient in our experiments.

\subsubsection{Application of GEV-MVDR BF to mixture enhancement}\label{sec:robustness-gev-mvdr}
As shown in \cite{GEV}, a GEV-MVDR BF can be viewed also as a realization of a Maximum Signal-to-Noise Ratio (MaxSNR) BF \cite{MPDR,maxSNR,heymann16icassp}. This property allows us to use it for mixture enhancement. We explain this in the following.

Regardless of whether the captured signal contains a single source or multiple sources, we can define a MaxSNR BF $\mathbf{w}_{t}^{\mbox{\scriptsize MaxSNR}}\in\mathbb{C}^M$ as one that maximizes the SNR of the signal, i.e., the power ratio of the desired signal $\mathbf{d}_t$ to the noise $\mathbf{v}_t$:
\begin{align}
    \mathbf{w}_{t}^{\mbox{\scriptsize MaxSNR}}&=\arg\max_{\mathbf{w}}\frac{E\{|\mathbf{w}^{\HT}\mathbf{d}_t|^2\}}{E\{|\mathbf{w}^{\HT}\mathbf{v}_t|^2\}},\\
    &=\arg\max_{\mathbf{w}}\frac{E\{|\mathbf{w}^{\HT}\mathbf{x}_t|^2\}}{E\{|\mathbf{w}^{\HT}\mathbf{v}_t|^2\}},\label{eq:maxsnr}
\end{align}
where we used $E\{|\mathbf{w}^{\HT}\mathbf{x}_t|^2\}=E\{|\mathbf{w}^{\HT}\mathbf{d}_t|^2\}+E\{|\mathbf{w}^{\HT}\mathbf{v}_t|^2\}$ to obtain the above second line.
The solution to Eq.~(\ref{eq:maxsnr}) is obtained based on GEV as
    $\mathbf{w}_{t}^{\mbox{\scriptsize MaxSNR}}=b^*\mathbf{u}_t$,
where $b$ is an indefinite constant, $(\cdot)^*$ is a complex conjugate, and $\mathbf{u}_t$ is from Eq.~(\ref{eq:maxeig}). Let $\mathbf{b}\in\mathbb{C}^M$ be a vector composed of $M$ different realizations of $b$ and let  $\mathbf{W}_t=\mathbf{u}_t\mathbf{b}^{\HT}$ be a matrix realization of the MaxSNR BF according to $\mathbf{b}$. Then, the MaxSNR BF with $\mathbf{b}$ defined below is identical to the GEV-MVDR BF in Eq.~(\ref{eq:gev-mvdr}).
\begin{align}
    \mathbf{W}_t^{\mbox{\scriptsize MaxSNR}}=\mathbf{u}_t\mathbf{b}^{\HT}\quad\mbox{where}\quad\mathbf{b}=\frac{\Phi_{\mathbf{v}}\mathbf{u}_{t}}{\mbox{tr}\{\mathbf{u}_{t}\mathbf{u}_{t}^{\HT}\Phi_{\mathbf{v}}\}}.\label{eq:maxsnr2}
\end{align}

The above analysis indicates that a GEV-MVDR BF in Eq.~(\ref{eq:gev-mvdr}) works as a MaxSNR BF even when the observed signal contains two or more sources. In other words, the GEV-MVDR BF may enhance the mixture of speech signals, i.e., $\mathbf{d}_t$ in Eq.~(\ref{eq:obs1}), in the MaxSNR sense regardless of the number of sources. 

\subsection{Two versions of Independent Component Extraction (ICE): ISCE and IMCE}
Next, we pick up Independent Component Extraction (ICE) as one that is applicable to mixture enhancement. ICE is a variation of a popular BSS technique, IVE \cite{Robin2019waspaa,ikeshita2021taslp,9616119}. Even without prior knowledge of ATFs of the sources, both IVE and ICE can extract a given number ($N \ge 1$) of sources based on an assumption that the sources are mutually independent. We here adopt ICE because our preliminary experiments showed the superiority of ICE over IVE for mixture enhancement. 
The difference between ICE and IVE is the source models they use. While IVE's model specifies the source's spectral characteristics over all frequencies, ICE's model specifies the characteristics separately at each frequency. We defined the ICE's model at each frequency by a complex Gaussian with a mean of 0 and a time-varying variance. The optimization algorithm for online ICE can be easily obtained from that of online IVE \cite{9616119}. 

We present two different usages of ICE, ISCE and IMCE, for mixture enhancement in the following.

\subsubsection{Application of ISCE to mixture enhancement}
When we set the number of sources to extract as $N=1$ for ICE, we call it Independent Single Component Extraction (ISCE). This paper uses ISCE for mixture enhancement by utilizing speech sparseness. With the speech sparseness, we assume that speakers are not simultaneously active at each TF point. This assumption is based on the spectral characteristics of speech and has proven effective for source separation purposes \cite{DUET,wang05book,souden10aslp}. With this assumption, we can simplify the mixture enhancement task to a single source extraction task at each TF point. Thus, we may use ISCE for mixture enhancement by simply applying ISCE to a mixture with quickly adapting ISCE over time frames. 

\subsubsection{Application of IMCE to mixture enhancement}
When we assume that the number of sources is given ($N\ge 2$) and use ICE to extract all the sources from a mixture, we call it Independent Multiple Component Extraction (IMCE). By remixing the extracted sources, we can obtain an enhanced mixture. Unlike the other methods, this method requires the prior knowledge of the number of sources. Thus, we use IMCE just for reference.  

\subsection{UR-MWF}
We can also use a version of conventional MWF, called UR-MWF \cite{URMWF}, for mixture enhancement. It is defined as a BF $\mathbf{W}_t^{\mbox{\scriptsize MWF}}$ that gives the minimum mean square error (MMSE) estimate of the desired mixture $\mathbf{d}_t$:
\begin{align}
    \mathbf{W}_t^{\mbox{\scriptsize MWF}}=\arg\min_{\mathbf W}E\{\|\mathbf{d}_t-\mathbf{W}^{\HT}\mathbf{x}_t\|_2^2\}.
\end{align}
Assuming $\Phi_{\mathbf{x},t}=E\{\mathbf{d}_t\mathbf{d}_t^{\HT}\}+\Phi_{\mathbf{v}}$, we obtain UR-MWF:
\begin{align}
    \mathbf{W}_t^{\mbox{\scriptsize UR-MWF}}=\Phi_{\mathbf{x},t}^{-1}(\Phi_{\mathbf{x},t}-\Phi_{\mathbf{v}}).\label{eq:URMWF}
\end{align}
Based on Eq.~(\ref{eq:URMWF}), UR-MWF can enhance a mixture given $\Phi_{\mathbf{x},t}$ and $\Phi_{\mathbf{v}}$ without knowing the number of sources or their ATFs.

\subsection{Comparison of computational time complexity}
Table~\ref{tab:my_label} shows the computational time complexity of the BFs and the Mod-PMWF.  All the BFs except for IMCE have the same complexity $O(M^2)$.  The complexity increases to $O(NM^2)$ for IMCE to extract $N$ sources.

\begin{table}[t]
    \centering
    \setlength{\tabcolsep}{3pt}
    \caption{Computational time complexity for adapting and applying a BF to each TF point. The complexity of Mod-PMWF, GEV-MVDR, and UR-MWF is estimated assuming that we recursively update $\Phi_{\mathbf{v}}$ and its inverse, e.g., during the speech absent periods, as in \cite{nakatani19_interspeech}.}
    \label{tab:my_label} 
    \begin{tabular}{ccccc}\toprule
    \rule[0mm]{0mm}{3mm}Mod-PMWF & GEV-MVDR & ISCE & IMCE & UR-MWF  \\\hline
    \rule[0mm]{0mm}{3mm}$O(M^2)$ & $O(M^2)$ & $O(M^2)$ & $O(NM^2)$ & $O(M^2)$ \\\bottomrule
    \end{tabular}
\end{table}

\section{Experiments}
\label{sec:experiment}
We evaluated our proposed Mod-PMWF in comparison with the other conventional BFs using signal level metrics and a MUSHRA listening test. 

\subsection{Dataset}
For the evaluation, we prepared a simulated noisy reverberant mixture dataset containing 100 samples of two speakers from the ATR digital speech database set B \cite{ATR_dataset}. To reverberate each speaker's utterance, we used room impulse responses (RIRs) measured with a cubic array of eight microphones in a room with a T60 of 380~ms. 
Each edge of the cubic array was 4 cm, and a microphone was equipped at each vertex of the array. We used 5 out of 8 microphones in the experiments. The two speakers were located 1 m from the array on opposite sides.  We then mixed two speech signals starting simultaneously with babble noise recorded by the same array in the same room. We created two versions of the dataset by setting the Reverberant speech Mixture to Noise Ratio (RMNR) to 20~dB and 10~dB. The sampling frequency was set at 16~kHz.

\subsection{Analysis condition}
Throughout the experiments, we set $\gamma=0$ for Mod-PMWF to evaluate its MVDR behavior in this paper. 
The forgetting factor $\beta$ for updating $\Phi_{\mathbf{x},t}$ in Eq.~(\ref{eq:scmupdate}) decides how fast the algorithm adapts to the change in the signals' spatial characteristics. We chose $\beta=0.99$ for UR-MWF, 0.96 for ISCE and IMCE, and 0.85 for Mod-PMWF and GEV-MVDR based on our preliminary experiments. We set the window length and shift at 20~ms and 5~ms.

In Sections~\ref{sec:results} and \ref{sec:mushra}, we assume that the interference SCMs $\Phi_{\mathbf{v}}$ can be estimated in advance. In concrete, we estimated them from a part of the measured babble noise not used for generating the mixtures. Then, we evaluated Mod-PMWF with no prior knowledge of $\Phi_{\mathbf{v}}$ in Section~\ref{sec:blind}, where we estimated $\Phi_{\mathbf{v}}$ by online processing from observed signals \cite{nakatani19_interspeech} using estimated speech presence probability \cite{soudenpsd}.


\subsection{Evaluation results using Signal Level Metrics}\label{sec:results}
We defined two signal level metrics for the evaluation \cite{Hansen98aneffective,metrics}: Segmental Desired to Interference Ratio (SegDIR) and Segmental Desired to Distortion Ratio (SegDDR). SegDIR was used for measuring the degree of interference reduction. SegDDR was used for measuring the degree of signal distortion introduced to the enhanced speech. 
Let $d_{m}(i)$ and $v_{m}(i)$ be the $i$th samples of the desired mixture and interference signal in the time domain at the $m$th microphone, and $\hat{d}_{m}(i)$ and $\hat{v}_{m}(i)$ be samples obtained by applying a BF to $d_{m}(i)$ and $v_{m}(i)$. To determine the desired signal $d_m(i)$, we set the duration of the early reflections at 50~ms. Then, the metrics are defined as:
\begin{subequations}
\begin{align*}
\mbox{SegDIR} &= \frac{10}{|\cal A|} \sum_{\tau\in{\cal A}} \log_{10}\frac{\sum_{m=1}^M\sum_{i=\tau\delta+1}^{\tau\delta+\delta} |\hat{d}_m(i)|^2}{\sum_{m=1}^M\sum_{i=\tau\delta+1}^{\tau\delta+\delta} |\hat{v}_m(i)|^2},\\
\mbox{SegDDR} &= \frac{10}{|\cal A|} \sum_{\tau\in{\cal A}} \log_{10}\frac{\sum_{m=1}^M\sum_{i=\tau\delta+1}^{\tau\delta+\delta} |d_m(i)|^2}{\sum_{m=1}^M\sum_{i=\tau\delta+1}^{\tau\delta+\delta} |\hat{d}_m(i) - d_m(i)|^2},
\end{align*}
\end{subequations}
where $\delta=800$ samples ($=50$~ms) is the segment length, $\tau$ is a segment index, ${\cal A}$ is a set of segments where at least one speech signal is active, and $|\cal A|$ is the number of segments in $\cal A$.


\begin{figure}
\centering
\includegraphics[width=\linewidth]{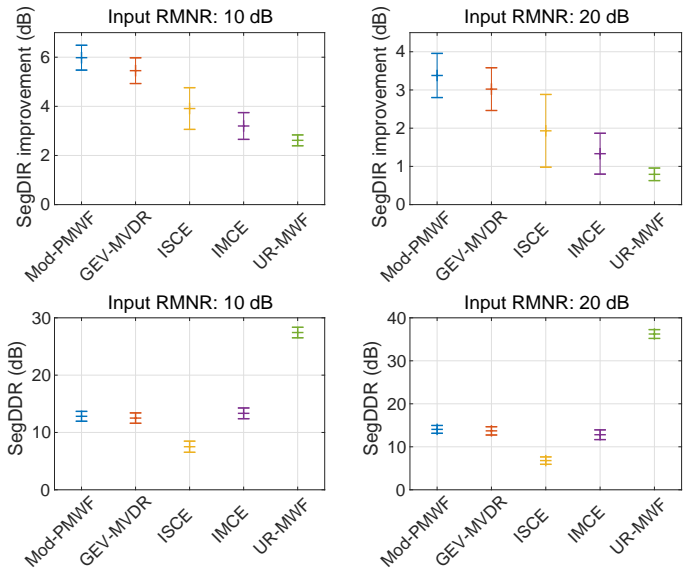}
\caption{SegDIR improvements and SegDDRs averaged over 100 samples. Error bars indicate standard deviation. Mod-MPDR, Gev-MVDR, and UR-MWF used interference SCMs separately estimated in advance.}
\label{fig:test}
\end{figure}

Figure~\ref{fig:test} shows the evaluation results. In the figure, we show SegDIR as its relative improvement from the captured signal and SegDDR as its absolute value. 

First, when we compare SegDIRs obtained by the BFs, the proposed Mod-PMWF was the best of all. In particular, Mod-PMWF significantly outperformed ISCE, IMCE, and UR-MWF.  Next, when we compare SegDDRs obtained by the BFs, Mod-PMWF was better than ISCM, and comparable with GEV-MVDR and IMCE. In contrast, UR-MWF achieved significantly better SegDDR than all the others, however, its improvement of SegDIR was minimal. 

We can draw the following implications from these results. 
\begin{enumerate}
\item Although Mod-PMWF and GEV-MVDR did not use prior knowledge of the number of sources, they achieved comparable SegDDRs with IMCE that uses the knowledge of the number of sources. This suggests that Mod-PMWF and GEV-MVDR effectively enhanced the mixture based on their unique mechanisms described in Sections~\ref{sec:ana2} and \ref{sec:robustness-gev-mvdr}. 
\item For improving SegDIR, the sparseness assumption used for ISCE was effective, to some extent, in comparison with IMCE, but insufficient in comparison with Mod-PMWF and GEV-MVDR. Also, the assumption introduced more speech distortion than the other BFs.
\item The inferior SegDIR obtained by IMCE is probably caused by remixing of two separated signals. The remixing sums up the interference that remained in each separated signal. In contrast, Mod-PMWF could effectively reduced the interference while enhancing the mixture. This is probably because Mod-PMWF adaptively controls the BF weights to enhance only active sources at each TF point. 
\end{enumerate}
In contrast, it is difficult to evaluate the effectiveness of UR-MWF in comparison with the other BFs based only on the above results.

To compare the proposed BF with the other BFs, including UR-MWF, more reliably, we conducted a subjective listening test.

\subsection{Evaluation Results with MUSHRA Listening Test}\label{sec:mushra}
We conducted a MUSHRA listening test \cite{MUSHRA} to evaluate the perceived speech quality. We picked up two speech mixtures composed of two male or two female speakers and tested them under each RMNR condition. Each sample had seven signals to rate, including the desired mixture as the (hidden) reference and the noisy reverberant observation as the anchor. Ten experienced listeners participated in the test. We asked them to concentrate on the overall quality of the audio including both the noise level and the degradation of speech.

\begin{figure}
\centering
\includegraphics[width=\linewidth]{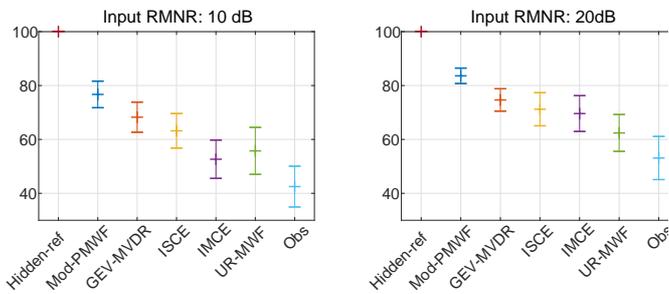}
\caption{MUSHRA test results. Error bars indicate 95\% confidence based on t-distribution test. Mod-MPDR, Gev-MVDR, and UR-MWF used interference SCMs separately estimated in advance.}\label{fig:mushra}
\end{figure}

Figure~\ref{fig:mushra} shows the result. The two sub-figures show the results under input RMNRs of 20~dB and 10~dB and both sub-figures exhibit almost the same tendency. First, the participants rated the high-quality reference as 100 points and test conditions between 40-80 points, confirming our choice of MUSHRA to be appropriate according to the ITU-R recommendation BS.1534. In addition, improvements obtained by all the BFs were statistically significant from the anchor, i.e., the noisy observation rated with the lowest scores. Thus our anchor can indicate how the methods compare to known audio quality levels. 

Next, when comparing results between BFs, the proposed Mod-PMWF significantly outperformed all the other BFs, including GEV-MVDR and UR-MWF. The result was better correlated with SegDIR than with SegDDR. One reason might be that the speech distortion is, to some extent, masked by the active speech and noise residual and thus not as annoying to the listeners as the noise residual.

\subsection{Evaluation with no prior knowledge of interference SCM}\label{sec:blind}
Finally, we evaluated Mod-PMWF without using prior knowledge of the interference SCM.  In this experiment, we recursively updated the interference SCM $\Phi_{\mathbf{v},t,f}$ at each TF point as
\begin{align}
\Phi_{\mathbf{v},t,f}&=\alpha'\Phi_{\mathbf{v},t-1,f}+(1-\alpha')\mathbf{x}_{t,f}\mathbf{x}_{t,f}^{H},\\
\alpha'&=\alpha+(1-\alpha)q_{t,f},
\end{align}
where $\alpha=0.9998$ is a forgetting factor and $q_{t,f}$ is a speech presence probability estimated recursively by maximum likelihood estimation \cite{soudenpsd}.

\begin{figure}
\centering
\includegraphics[width=\linewidth]{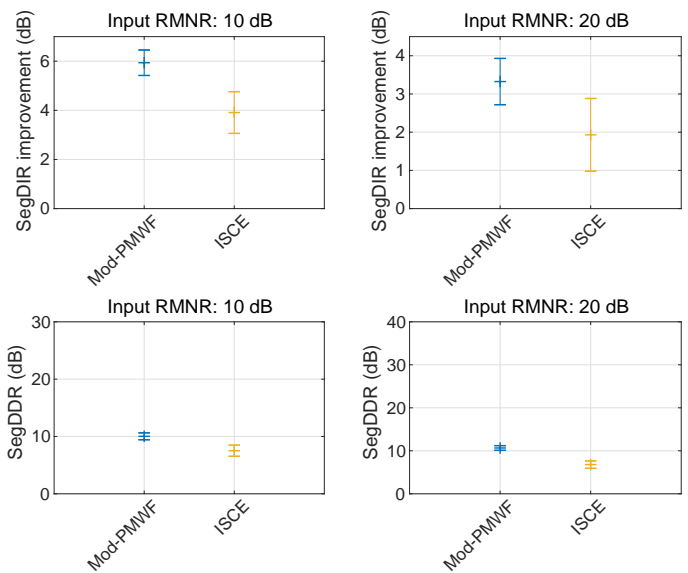}\\
{(a) SegDIR improvements and SegDDRs averaged over 100 samples. Error bars indicate standard deviation.\vspace{3mm}}
\label{fig:test2}
\includegraphics[width=\linewidth]{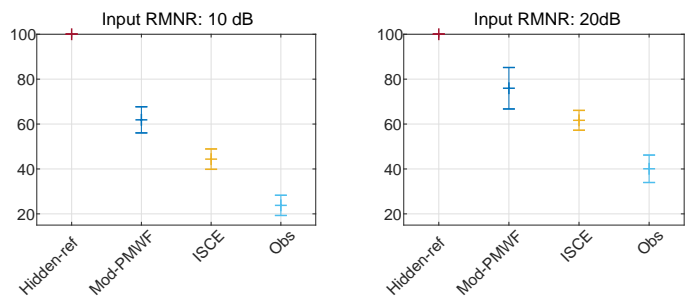}\\
{(b) MUSHRA test results. Error bars indicate 95\% confidence based on t-distribution test.\vspace{3mm}} 
\caption{Evaluation with no prior knowledge of interference SCM. The interference SCM was estimated by online processing for Mod-PMWF. }\label{fig:no prior}
\end{figure}

Figures~\ref{fig:no prior} (a) and (b) show SegDIR improvements, SegDDRs, and MUSHRA test results.  In this experiment, we selected two methods, Mod-PMWF based on online interference SCM estimation and ISCE based on BSS, mainly to elaborate reliability of the MUSHRA test. For the MUSHRA test, we also added a mixture sample composed of male and female speakers. Totally we had 3 samples for each RMNR condition. Eight experienced listeners participated in this test.

Mod-PMWF was significantly better than ISCE in terms of all the metrics. The overall tendency of the results in the figures was almost identical to that in Figs.~\ref{fig:test} and \ref{fig:mushra} except that SegDDRs obtained by Mod-PMWF were slightly lower than Fig.~\ref{fig:test}. These results show that Mod-PMWF effectively performed mixture enhancement even when no prior knowledge on the interference SCMs was available.

\section{Conclusion}
\label{sec:conclusion}
This paper proposed a Mod-PMWF for low-latency online speech mixture enhancement in noisy reverberant environments. We showed mathematically that Mod-PMWF is equivalent to a weighted sum of MVDR BFs, where each MVDR BF can preserve each source included in the captured signal and the time-varying weights quickly adapt at each time frame to put larger weights on sources with larger powers. We can estimate the Mod-PMWF by low-latency online processing without relying on prior knowledge of the number of sources or the ATFs from the sources to microphones. We discussed that approximating the SCM of the desired mixture by the SCM of the captured signal for Mod-PMWF is advantageous to make it accurately preserve the speech mixture although it slightly reduces the interference reduction effect. 
We verified the effectiveness of the Mod-PMWF for mixture enhancement in terms of interference reduction ratios and subjective speech quality tests. The Mod-PMWF outperformed several conventional BFs, including GEV-MVDR BF that performs mixture enhancement in a MaxSNR sense, a BSS technique ISCE utilizing the speech sparseness, another BSS technique IMCE with prior knowledge of the number of sources, and UR-MWF that gives the MMSE estimates of the desired mixtures.

\footnotesize
\printbibliography

\end{document}